\documentclass[preprint]{aastex62}

\newcommand{\speed}[1]{#1 km~s${}^{-1}$}
\newcommand{\Mx}[1]{#1 Mx~s${}^{-1}$}
\newcommand{\accel}[1]{#1 m~s${}^{-2}$}

\usepackage{lineno}
\shorttitle{macrospicules}
\shortauthors{Duan et al.}

\received{... ..., ....}
\revised{... ..., ....}
\accepted{... ..., ....}
\submitjournal{ApJL}

\begin{document}
\title{Macrospicules and Their Connection to Magnetic Reconnection in the Lower Atmosphere}
\correspondingauthor{Yuandeng Shen}
\email{ydshen@ynao.ac.cn}

\author[0000-0001-9491-699X]{Yadan Duan}
\affil{Yunnan Observatories, Chinese Academy of Sciences, Kunming, 650216, China}
\affil{State Key Laboratory of Space Weather, Chinese Academy of Sciences, Beijing 100190, China}
\affil{University of Chinese Academy of Sciences, Beijing 100049, China}
\affil{Yunnan Key Laboratory of Solar Physics and Space Science, Kunming, 650216, China}

\author[0000-0001-9493-4418]{Yuandeng Shen}
\affil{Yunnan Observatories, Chinese Academy of Sciences, Kunming, 650216, China}
\affil{State Key Laboratory of Space Weather, Chinese Academy of Sciences, Beijing 100190, China}
\affil{University of Chinese Academy of Sciences, Beijing 100049, China}
\affil{Yunnan Key Laboratory of Solar Physics and Space Science, Kunming, 650216, China}
\author[0000-0001-7866-4358]{Hechao Chen}
\affil{School of Physics and Astronomy, Yunnan University, Kunming 650500, People’s Republic of China}
\author{Zehao Tang}
\affil{Yunnan Observatories, Chinese Academy of Sciences, Kunming, 650216, China}
\affil{University of Chinese Academy of Sciences, Beijing 100049, China}
\author{Chenrui Zhou}
\affil{Yunnan Observatories, Chinese Academy of Sciences, Kunming, 650216, China}
\affil{University of Chinese Academy of Sciences, Beijing 100049, China}
\author{Xinping Zhou}
\affiliation{Sichuan Normal University, College of Physics and Electronic Engineering, Chengdu 610068, People’s Republic of China}
\author{Song Tan}
\affil{Yunnan Observatories, Chinese Academy of Sciences, Kunming, 650216, China}
\affil{University of Chinese Academy of Sciences, Beijing 100049, China}

\begin{abstract}
Solar macrospicules are beam-like cool plasma ejections of size in-between spicules and coronal jets, which can elucidate potential connections between plasma jetting activity at different scales. With high-resolution observations from the {\em New Vacuum Solar Telescope} and {\em Solar Dynamic Observatory}, we investigate the origin of five groups of recurrent active-region macrospicules. Before the launch of each macrospicule, we detect a compact bright patch (BP) at its base where a newly emerging dipole contacts and cancel with the pre-existing ambient field. The spectral diagnosis from the {\em Interface Region Imaging Spectrograph} at one of BPs reveals signatures of reconnection at the lower atmosphere. Multiwavelength imaging of these BPs show that they mainly occur at the rising phase of the flux emergence and slowly ascend from the lower to the upper chromosphere. Remarkable macrospicules occur and fade out once the BPs appear and decay from the AIA 304 \AA~images, respectively. We suggest that these macrospicules and related BPs form in a common reconnection process, in which the increasing reconnection height between the emerging dipole and the ambient field results in the observed variations from BPs to macrospicules. Interestingly, most macrospicules show similar characteristics to larger-scale coronal jets and/or smaller-scale spicules, i.e., the rotating motions, the presence of minifilaments and BPs before the eruptions, and magnetic flux emergence and cancellation. We conclude that the formation mechanism of macrospicules should be the same as spicules and coronal jets, i.e., solar jetting phenomena at different scales share the same physical mechanism in association with magnetic reconnection.
\end{abstract}


\section{Introduction} \label{sec:intro}
The solar atmosphere is full of transient and dynamic plasma ejections. These fascinating plasma jetting activities shoot out from the solar surface into higher atmospheric layers at various size scales and appear to occur in all solar regions (but seemingly avoiding sunspots), which are regarded to be one of the most significant sources for heating coronal plasma and accelerating solar wind because they constantly replenish mass and energy fluxes into the upper atmosphere. According to their size scales, these jetting activities are usually divided into spicules, coronal jets, and macrospicules in-between the former two types of plasma ejections in size. Spicules are a component of the chromosphere, which can be characterized as rapid blue- or red-shifted excursions (RBEs or RREs) in on-disk spectroscopic observations \citep[e.g.,][]{2008ApJ...679L.167L,2009ApJ...705..272R}. They are thin (the widths in the range of less than an arcsecond) structures that can be detected in H$\alpha$, Ca~{\sc{ii}}, and He~{\sc{ii}} images  \citep[e.g.,][]{2014ApJ...792L..15P,2015ApJ...806..170S}, and their lengths most below 5000 km. In the light of the proposal by \citet{2007PASJ...59S.655D}, spicules could be classified as type I (velocity $\sim$ 15$-$40 $\speed$; lifetime $\sim$ 3$-$10 minutes; ) and type II (velocity $\sim$30$-$110 $\speed$; lifetime $<$100 s). Compared to spicules, coronal jets occur at a relatively larger spatial scale, with lengths ranging from $\sim10^{4}-10^{5}$ km, widths between $\sim$ 5000 and 10$^5$ km, a mean velocity of 200 $\speed$, and a lifetime of $\sim$ 10 minutes \citep[e.g.,][]{1996PASJ...48..123S,2007PASJ...59S.771S}. Based on different observing wavelengths, they can be classified as extreme ultraviolet (EUV) jets, and X-ray jets.

Over the past three decades, extensive observational studies and numerical simulations have studied the triggering and driving mechanisms of coronal jets. It has been widely acknowledged that magnetic reconnection is the primary energy release mechanism in coronal jets \citep[e.g.,][]{2016SSRv..201....1R,2021RSPSA.47700217S,2021GMS...258..221S}. Some potential theoretical models that are related to the generation and dynamics of spicules including shock waves \citep{2004Natur.430..536D}, Alfv$\acute{e}$n  waves \citep{2015ApJ...812...71C,2017ApJ...848...38I}, amplified magnetic tension \citep{2017ApJ...847...36M} and magnetic reconnection \citep{2011A&A...535A..95D,2019Sci...366..890S}. However, so far the generation mechanism of solar spicules is still an open question. The idea of magnetic reconnection for the generation of spicules has gained increasing attention. For instance, using 2.5-dimensional MHD numerical simulations, \citet{2001ApJ...546L..73T} investigated photospheric magnetic reconnection induced by convective intensification of solar surface magnetic fields. In their model, the upward-propagating slow mode MHD waves were generated as a result of the photospheric reconnection, and the energy of these waves were sufficient to produce spicules. Thus, they suggested that photospheric magnetic reconnection might be one of the causes of solar spicules. With early ground-based solar telescopes, a few observations gave some clues that the interaction between emerging flux and ambient field might be responsible for the occurrence of spicules \citep[e.g.,][]{2013ApJ...767...17Y, 2015ApJ...799..219D}, but their relatively poor seeing and low time cadence impeded a deeper study of the spatio-temporal association between spicules and photospheric magnetic field evolution. Using high-resolution H$\alpha$ and vector magnetic field observations from the Goode Solar Telescope, \cite{2019Sci...366..890S} recently found that enhanced spicular activities are possibly associated with the emergence of minor magnetic flux elements and/or subsequent cancellation with an opposite-polarity magnetic flux patch. This observational signal is the same as what has been identified in large-scale coronal jets driven by magnetic reconnection, and it might imply that at least some spicules might not be too different from these reconnection-driven coronal jets but only different in size. If this is the case, jetting activities on different size scales might both be correlated to magnetic reconnection. However, resolving the detailed triggering and driving processes of spicules at their bases is still difficult, since these key processes occur at a small scale close to or below the highest spatial resolution of most current instruments.
%how reconnection takes part in and what role it plays in the triggering spicules remains a mystery

Macrospicules are described as giant spicules detected in the lower solar atmosphere \citep{1975ApJ...197L.133B,2000A&A...360..351W}. Previous works have studied their basic physical properties, such as lifetime (3$-$45 minutes), width (3$-$16 Mm), length (7$-$70 Mm), and maximum velocity (10$-$150 $\speed$) \citep[e.g.,][]{1975ApJ...197L.133B,2015ApJ...808..135B,2017ApJ...835...47K,2019ApJ...871..230L}.
\citet{1998SoPh..182..333P} observed simultaneous blueshifted and redshifted emissions on each side of the macrospicules' axis, which can be interpreted as the presence of rotating motion around the spires \citep[e.g.,][]{2010A&A...510L...1K,2011A&A...532L...9C}. \citet{1977ApJ...218..286M} and \citet{1979SoPh...61..283L} identified some macrospicules with mini-filament eruptions, whereas \citet{2000ApJ...530.1071W} presented that macrospicules and eruptive mini-filament were two different forms of activity. Furthermore, \citet{2013ApJ...770L...3K} studied a small-scale twisted flux tube that undergoes internal reconnection and triggers a macrospicule and an associated coronal jet. Combined with two-dimensional numerical simulation, they showed that the internal reconnection further generates a velocity pulse that steepens into slow-shock waves propagating upward and triggers the macrospicule and the jet. In addition, very small-scale bright patches, the possible proxy of magnetic reconnection, appearing below macrospicules have been reported in a few observations \citep{1977ApJ...218..286M,1997SoPh..175..457P, 2000SoPh..194...59Z,2017ApJ...835...47K,2019ApJ...871..230L}. However, to fully clarify how reconnection takes part in and what role it plays in the formation of macrospicules, new multi-wavelength and high-resolution observations of macrospicules are still highly needed.

With joint observations from the New Vacuum Solar Telescope \citep[NVST,][]{2014RAA....14..705L, 2016NewA...49....8X}, Interface Region Imaging Spectrograph \cite[IRIS,][]{2014SoPh..289.2733D}, the Atmospheric Imaging Assembly \cite[AIA,][]{2012SoPh..275...17L} and the Helioseismic and Magnetic Imager \cite[HMI,][]{2014SoPh..289.3483H} onboard the Solar Dynamic Observatory, we investigated the triggering of five groups of recurrent active-region macrospicules and the associated signal observed in the lower atmosphere. The present study provides new evidence to support that macrospicules are driven by magnetic reconnection in the low solar atmosphere, which also suggests the potential similarity among spicules, macrospicules, and coronal jets occurring at different size  scales.

\section{Observations And Data Analysis} \label{sec:instr}
The NVST is desired for studying fine structures in the chromosphere of the Sun with a temporal resolution of 45 s and a pixel resolution of 0\arcsec.165 in the H$\alpha$ channel. On November 11, 2020, NVST captured four groups of recurrent macrospicules in an active region (AR 12781) near the center of the solar disk from 01:46 UT to 09:00 UT, enabling us to study their triggering processes.
Meanwhile, the IRIS point was (319\arcsec, -429\arcsec), which was close to the disk center and recorded another recurrent macrospicule in the same active region.
\par
The data used in this study were H$\alpha$ line center at 6562.8 \AA~and H$\alpha$ $\pm$0.6 \AA~off-band images taken by the NVST. 
Based on the H$\alpha$ $\pm$0.6 \AA~off-band images, Doppler proxy maps can be constructed following this formula: $D$ = ($I_{red}$ $-$ $I_{blue}$)/($I_{red}$ $+$ $I_{blue}$), where $D$ represents the Doppler proxy index, and $I_{red}$($I_{blue}$) are the emission intensities of the blue (red) wing \citep[e.g.,][]{2020ApJ...902....8C}. Meanwhile, the IRIS level-2 slit-jaw image in the filter of Si {\sc{iv}} 1400 \AA~(0.08 MK) and the Mg {\sc{ii}} k 2796 \AA~(0.01 MK), and raster scan spectral data were used.
The photospheric magnetic field and the observational images of the five groups of recurrent active-region macrospicules were respectively taken by the AIA and HMI. The AIA EUV (UV) images have a cadence of 12s (24s) and both a pixel resolution of 0\arcsec.6. The cadence and pixel size of HMI images are 45s and 0\arcsec.5, and the measuring precision of the HMI light-of-sight magnetogram is $\pm$ 10 G. All imaging data were co-aligned by matching commonly observed features in AIA 304 \AA~(1600 \AA) and NVST line center 6562.8 \AA\ (SJI 1400 \AA) images.

\section{RESULTS} \label{sec:OBS}
During 2020 November 10 to 11, five groups of recurrent macrospicules (we call them C1-C5 hereafter) were observed in on-disk active region 12781, and their location was shown in Figure 1 (a). Four of them were recorded by the NVST and one was recorded by the IRIS. As shown in Figure 1, they were all identified by the chromosphere line and occurred at a similar mixed-polarity region where a newly-emerging dipole intruded into an opposite-polarity magnetic flux region. For each case, the newly emerging dipole was overlaid on the corresponding IRIS/SJI 2796 \AA~and NVST H$\alpha$ -0.6 \AA~images (see red contours in figure 1 (c1)$-$(c5)). As shown in the online animation, it can be observed more straightforwardly that all macrospicules originated from locations where negative emerging polarities intruded into the dominated pre-existing positive magnetic flux (see also Figure 1 (b1-b5)). Such a magnetic field configuration provides a favorable place for the occurrence of magnetic reconnection and is analogous to the fan-spine magnetic topology of coronal jets in three-dimension. In SJI 2796 \AA~and NVST H$\alpha$ -0.6 \AA~images, we detected a compact bright patch (BP) at or near the base of each macrospicule (see the online animation). Meanwhile, a faint inverted Y-shaped structure can be discerned in one of these events ( see Figure 1 (c3)). These observational characteristics are suggestive signatures of magnetic reconnection, indicating that all these macrospicules may originate from a reconnection process occurring at their base due to the interaction between the dominant positive ambient fields and the negative flux of emerging dipoles. Moreover, four of these macrospicules exhibited simultaneous blueshift and redshift emissions on both sides of the main axes (see Figure 1 (d2)$-$(d5)). Such a doppler pattern suggests the rotating motion of the macrospicules \citep[e.g.,][]{1998SoPh..182..333P,2010A&A...510L...1K,2011A&A...532L...9C} similar to those identified in many rotating coronal jets \citep[e.g.,][]{2011ApJ...735L..43S, 2013ApJ...769..134M, 2021RSPSA.47700217S, 2016SSRv..201....1R}.
\par
In Figure 2, we show a close-up view of the morphological evolution of case C3 in different solar atmosphere layers and the accompanied small-scale filament. This case started with a BP that was nearly invisible at 06:34 UT and then showed up in the NVST images at 06:51 UT. However, this feature was absent from AIA 304 and 211 \AA~images. This result indicates that the BP likely occurred in the low or middle chromosphere. Subsequently, this BP can be detected in a higher atmospheric layer from the lower chromosphere up to the transient region. In this course, the macrospicule appeared in the AIA 304 \AA~, and 211 \AA~ passbands, and it was shown as a linear absorption feature in various wavelength images at 07:36 UT. At about 07:42 UT and 07:45 UT, the macrospicule showed different morphologies in H$\alpha$ wing, H$\alpha$ line center, and AIA images (see the contour in Figure 2). Afterwards, when the macrospicule faded out of the H$\alpha$ images, it was still visible in the AIA 304 \AA~, and 211 \AA~images (see 07:49 UT). These results suggest that the macrospicule was composed of multiple plasma flows at different temperatures. 

\par
Interestingly, an accompanied small-scale filament is observed at the base of the recurrent macrospicules C3 (see the last row in Figure 2 ). The small-scale filament was shown a semicircular shape with a length of 3$-$4 Mm. The length of this small-scale filament is $\sim$ 2 times smaller than those reported in \citet{2015Natur.523..437S} and \citet{2020ApJ...902....8C}  ($\sim$ 8 Mm), therefore, we would like to call it small minifilament. According to the time sequence images of the high-resolution NVST H$\alpha$ line center, the small minifilament formed between 06:00 UT and 06:34 UT, and it was accompanied by the negative magnetic flux emergence (see the red contours of the last row in Figure 2). Note that we checked our five groups of recurrent macrospicules and found that only two groups had small minifilaments. One group is shown here, and the other one is recorded in Table 1. With the continuous flux emergence and canceling with ambient opposite-polarity magnetic fields, the small minifilament erupted to form the macrospicule (at 06:53 UT and 07:36 UT see Figure 2 ). This result might suggest that the upward cold plasma flow of the macrospicule was the erupting small minifilament, similar to those identified in coronal jets \citep[e.g.,][]{2012ApJ...745..164S,2017ApJ...851...67S,2015Natur.523..437S,2017ApJ...835...35H,2019ApJ...881..132D,2020ApJ...902....8C}.
\par
For all macrospicules observed in the present study, we find that each macrospicule was associated with the first emergence and then cancellation of a negative polarity, and a small BP appeared around the base of the macrospicule, which experienced an ascending phase from the lower chromosphere to the transient region, followed by a descending phase in which the BP sequentially disappeared from the low corona to the lower chromosphere (see Table 1). Taking macrospicule C3 as an example, we show the radiation variations of the macrospicule base in different wavelength images in Figure 3 (a). At about 06:30 UT before the event, the magnetic source region in the HMI magnetogram was a region of positive magnetic flux, and nothing could be observed in the Ha, AIA 1600 \AA~, 304 \AA~, and 171 \AA~ images. At about 06:34 UT, when a negative polarity emerged in the magnetic source region, a compact BP appeared in the H$\alpha$ -0.6 \AA images. Afterwards, the negative flux of the emerging polarity increased its area to the maximum at 07:14 UT, and then it rapidly cancelled with the pre-existing dominant positive flux.  During this course, the compact BP sequentially appears in the AIA 1600 \AA~, 304 \AA~ and 171 \AA~ images. At about 07:36 UT, the BP reached its maximum brightness at 304 \AA~ and 171 \AA~ images, and it could be observed simultaneously in all H$\alpha$, UV, and EUV passbands (see the fifth row of Figure 3). Note that the AIA 171 \AA~ channel contains contributions from cool line and hot lines. The contribution from the cool line is relatively faint, and the observed BP is also faint relative to the background intensity. This result suggests that when the BP reaches its maximum brightness at 171 \AA~ channel, it may occur in the transient region rather than the corona. 
\par
After 07:14:40 UT, the emerging negative polarity underwent a decay phase and finally disappeared from the HMI magnetogram at about 07:58:10 UT. In the meantime, the BP retreated sequentially from the AIA 171 \AA~, 304 \AA~, 1600 \AA~ and H$\alpha$ images. The enhanced radiation of the BP across wavelengths from H$\alpha$ to EUV passbands and its unique temporal evolution indicates that before the launch of the macrospicule, a strong plasma heating had happened in the chromosphere and then it gradually ascended to the transient region. It is worth noting that we tracked the centroids of the macrospicule's footpoints, and we found that the bright path (namely the northeast footpoint) showed a moving trend from southwest to northeast over time, while the southwest footpoint displayed a moving trend from northeast to southwest (see Figure 3 (b) and the animation). This spreading of the macrospicule's footpoints looks like the expending of flare ribbons observed in the standard solar flare model, which might indicate that the height of the reconnection site between the emerging loops and ambient open fields increases dynamically. Combining with the surface area of the negative flux that first increased gradually and then decreased, we tend to believe that these progressively evolving from chromospheric to UV emission of BPs most likely suggests the rising of the post-flare loop and the reconnection site. During the emerging process of the negative polarity (before 07:10 UT, see the red curve in Figure 4 (c)), there was no obvious plasma ejection in EUV channel (also see Figure 4 (c)). However, it is interesting to note that the macrospicule can be obvious observed in all passbands when the emerging negative polarity starts to decay (after 07:15 UT). In the rest of the studied cases, a similar evolution process can be observed and their detailed information is listed in Table 1. These observational results suggest that the BP can be regarded as a flare loop (see the red curves in Figure 3 (c)) which is produced by the reconnection between the emerging magnetic flux and the ambient unipolar fields. The sequential appearance of the BP from chromospheric to transient region might suggest the rising of the post-flare-loop and the reconnection site (as marked by the black cross sign in Figure 3 (c)). Moreover, the occurrence of macrospicules (a typical macrospicule is shown in Figure 3 (b)) can be interpreted by the classical emergence-cancellation-driven reconnection model.

\par
To get insight into the kinematics of the studied macrospicules, we made space-time diagrams by using AIA 193~\AA~ images, which best showed the ejecting trajectories of the macrospicules in four events and provided a more persistent observation. The slit cuts used to make the space-time diagrams are along the main axis of the macrospicules. The results are shown in Figure 4 (a)$-$(d), and the corresponding magnetic flux of the negative polarity is overlaid in each time-space diagram as a red curve. Here, we only show the time-space diagrams of events C1$-$C3 and C5 (see Figure 4 (a)$-$(d)). Since the case C4 exhibited a complex lateral slipping motion, so that it is hard to trace the trajectory of the recurrent macrospicules, we do not show the time-space diagram here. In the time-space diagrams, each macrospicule is shown as a dark parabolic path. As indicated by the black curves, for each event, one can identify many recurrent macrospicules. As observed in direct images (see Figure 4 (a)$-$(d)), with the emergence and cancellation of magnetic flux, the frequent occurrence of compact brightening can be identified at the base of the macrospicules. The interesting thing is that the upward ejecting macrospicules always fade out with the disappearance of the associated brightening. These features further reinforce that these macrospicules are all reconnection-driven activities, and they are highly associated with obvious flux emergence and cancellation. In addition, this also implies that the ejecting macrospicules plasma lost their driver due to the weakening or stopping of magnetic reconnection.

\par
We estimated the decelerations and maximum velocities of 40 well-defined trajectories through parabolic fittings from the four space-time diagrams. The results are shown in Figure 4 (e) and Table 1. The ejecting height of the observed macrospicules is in the range of 15$-$34 Mm, and the maximum width is about 1.31$-$5.66 Mm. These parameters are basically similar to those reported in previous literature \citep[e.g.,][]{1975ApJ...197L.133B,2015ApJ...808..135B,2017ApJ...835...47K,2019ApJ...871..230L}, suggesting that all these jet-like features are typical macrospicules. They showed a succession of upward and downward plasma flows following a parabolic path (see Figure 4 (a)$-$(d)). Moreover, we find a good linear correlation between the deceleration and maximum velocity (see Figure 4 (e)). This is quite similar to the nature of upward propagating magnetohydrodynamic (MHD) shock waves in numerical simulations of shock wave-driven dynamic fibrils and chromospheric jets \citep{2006ApJ...647L..73H, 2007ApJ...655..624D, 2007ApJ...666.1277H}. The deceleration is generally between 60$-$400$\accel$. The maximum velocity is mainly in the range of 20$-$120$\speed$, which is close to the sound speed in the transition region. In addition, the period of these recurrent macrospicules is about 10 minutes, longer than the MHD shock periods (i.e., 3$-$5 minutes; generated by the p-mode leaks from the photosphere) of smaller-scale chromospheric jets, spicules, and dynamic fibrils \citep[e.g.,][]{2004Natur.430..536D, 2007ApJ...655..624D,2006ApJ...647L..73H,2007ApJ...666.1277H}. These results reinforce the scenario that macrospicules are driven by magnetic reconnection.

\par
Case C1 was well observed by the IRIS raster scan (see Figure 5 (a), (b1), and (b2)), which provides us a good opportunity to study the associated spectral variation of the BP around the base of the macrospicule. As shown in Figure 5 (b1) and (b2), the macrospicule shows as an elongated dark feature in the Mg~{\sc{ii}} 2796 \AA~ SJI images (outlined by the dashed yellow lines), and the BP can well be observed in both Mg~{\sc{ii}} 2796 \AA~ and Si~{\sc{iv}} 1400 \AA~ SJI images. Compared with the reference spectral profiles (the black lines) in a quiescent region, the spectral profiles (red lines) at the BP are characterized by an apparent enhancement and broadening in both C~{\sc{ii}} and Si~{\sc{iv}} lines, without obvious superposition of chromospheric absorption lines (e.g., Ni~{\sc{ii}} 1393.33 \AA\ and 1335.20~\AA). These spectral signatures at the BP are quite similar to the UV bursts that are not connected to Ellerman bombs (EBs) \citep[e.g.,][]{2016ApJ...824...96T,2018ApJ...854..174T,2019STP.....5b..58H}.
The line wings of Si~{\sc{iv}} 1393.755~\AA~ and 1402.770~\AA~ reveal an obvious enhancement which possibly supports the presence of an associated outflow \citep[e.g.,][]{1997Natur.386..811I}, and the dramatic increase in the intensity of Si~{\sc{iv}}, as well as C~{\sc{ii}}, is caused by the heating of the cool plasma via magnetic reconnection occurring in the upper chromosphere \citep[e.g.,][]{2015ApJ...811...48Y}. Consistent with the NVST off-band observations, these results also suggest that the reconnection-induced BP observed at the macrospicule base took place in the upper chromosphere. 
To our knowledge, this is the first evidence that macrospicules are connected to solar UV bursts, and the  two types of small-scale activities likely form at different heights but originate from a common reconnection process \citep[e.g.,][]{2019ApJ...875L..30C}.

\section{Summary and Discussion} \label{sec:summ}
In this work, we employed NVST, IRIS, and SDO observations to investigate five groups of recurrent on-disk macrospicules and their corresponding responses in the lower solar atmosphere. Our study not only provides new constraints to the modeling of macrospicules but also reveals a potential similarity between solar jet-like activities at different size scales. 
Our main findings can be summarized below:
  
\begin{description}
\item[(i)]{All macrospicules were launched from the locations where newly emerging polarities interacted and canceled with the pre-existing ambient field of opposite polarities. In terms of temporal relationship, all the macrospicules tend to occur in the rising phase of the flux emergence and are closely related to an emergence-cancellation-reconnection process. Before the launch of the macrospicules, small minifilaments can be seen at the base of two groups of recurrent macrospicules.}

\item[(ii)]{At the base of each macrospicule, we detected compact BPs in EUV/UV and H$\alpha$ passbands, as well as H$\alpha$ off-bands. The H$\alpha$, AIA 1600 \AA, 304 \AA, 171 \AA~observations revealed that these compact BPs first formed in the lower chromosphere and then successively extended to the upper chromosphere and even the transient region. Moreover, these BPs tend to appear before the launch of the corresponding macrospicule and fade out around the drop of macrospicules.}

\item[(iv)]{The IRIS spectral diagnosis at the BP of one macrospicule revealed telltale signatures of magnetic reconnection in the lower atmosphere: an apparent enhancement and broadening in both C~{\sc{ii}} and Si~{\sc{iv}} lines but without obvious superposition of chromospheric absorption lines (e.g., Ni~{\sc{ii}} 1393.33 and 1335.20~\AA). Such unique spectral signals at the base of macrospicules suggest that this BP can be characterized by EB-unrelated UV bursts.}

\item[(iv)]{The NVST H$\alpha$ wing revealed a simultaneous appearance of blueshift and redshift emissions on each side of the spire of the macrospicules, indicating their rotating motion}.

\item[(v)]{The macrospicules showed a succession of distinct upward and downward motions, and their acceleration and maximum velocity exhibited a linear correlation. The period of the studied macrospicules (10 minutes) was much longer than those reported in smaller-scale chromospheric jets, spicules, and dynamic fibrils (3$-$5 minutes). The maximum velocity of the macrospicules (20-120 km/s) was close to the local sound speed of transition region.}
\end{description}

%macrospicules and jets
Magnetic reconnection has become a popular explanation for the triggering of coronal jets \citep[e.g.,][]{1996PASJ...48..353Y, 2016A&A...596A..36P, 2021RSPSA.47700217S}, since it has been supported by many observations \citep[e.g.,][]{2012ApJ...745..164S, 2019ApJ...883..104S, 2019ApJ...885L..11S, 2021ApJ...911...33C, 2019ApJ...881..132D, 2022ApJ...926L..39D}.  Using magnetohydrodynamic (MHD) numerical simulations, \citet{1995Natur.375...42Y,1996PASJ...48..353Y} well reproduced the observed characteristics of X-ray jets as a result of magnetic reconnection between the emerging flux and the pre-existing ambient coronal field. \citet{2007Sci...318.1591S} proposed that reconnection can be expected to occur at a relatively small spatial scale if a tiny dipole collides with an ambient field of opposite polarity, which can effectively result in Ca~{\sc{ii}} jets in the chromosphere. Meanwhile, they also found that numerous tiny jet-like structures were ejected from bright points and exhibited typical jet-like structures, whose footpoints were not simple bright points but an inverted Y-shape. \citet{2009SoPh..259...87N} also found that the appearances of EUV jets were often related to bright points at their bases; occasionally, sympathetic jets can be observed in adjacent coronal bright points \citep[e.g.,][]{2021ApJ...912L..15T}. Recently, similar bright points have also been reported at the bases of microjets \citep[e.g.,][]{2021ApJ...918L..20H,2022ApJ...928..153H}. This appears to mean that the morphological characteristics of the heated plasma at the base of jet-like structures may be caused by reconnection occurring at various heights, respectively. We also notice that similar bright patches beneath macrospicules \citep[e.g.,][]{1977ApJ...218..286M,1997SoPh..175..457P, 2000SoPh..194...59Z, 2017ApJ...835...47K} have been reported in a few previous works. In comparison, our current work provides a higher-resolution observation with much more rich physical information. In our study, we examined five groups of recurrent macrospicules, and a faint inverted Y-shaped structure was observed at the base of case C3. All the studied macrospicules were associated with BPs close to their bases, and we tend to believe that they were all in an inverted Y-shaped structure. The difference in morphology may be due to the difference in reconnection height (as suggested by \citet{2007Sci...318.1591S}) or influenced by the projection effect. 
Our results support the following scenario, as shown in Figure 3 (c): A small-scale dipolar emerges below the photosphere and interacts with the pre-existing ambient unipolar fields, twisted fields or small minifilament form at the base of the macrospicules. During this process, magnetic reconnection is expected to occur at the interface between opposite-polarity magnetic fields. As a result, a small flaring loop (BP) and a group of open fields would be newly produced. Meanwhile, due to the energy release at the reconnection site, the localized plasma would be heated and ejected outward along the reconnected field lines, forming the macrospicules. The BPs in our work might be the same as the brightening points in jet observations, though the specific morphological characteristics cannot be spatially resolved due to their lower reconnection height and the limitation of our observations.

\par
In the present study, the observed transition from compact BPs to cool plasma ejections at the base of macrospicules is the key to understanding the triggering of macrospicules. We conjecture that such a transition results from the magnetic reconnection between the emerging dipole and the ambient field, which has an increasing reconnection height as evidenced in previous multi-wavelength observations and the spreading of bright footpoints at the macrospicules. As revealed by the recent theory and simulations \citep[e.g.,][]{2018ApJ...862L..24P, 2019A&A...628A...8P, 2020ApJ...891...52S}, the solar atmosphere's responses to reconnection (including energy partition and dissipation) largely depend on the height of the reconnection location. In our case, reconnection may occur in a vertical/oblique current sheet through the interaction between the emerging dipole and the ambient unipolar fields. In the early phase, the newly twisted fields or small minifilaments would appear at the base of macrospicules during the period of dipole emergence. The newly-emerged dipole would cancel with the ambient fields at a relatively lower height at the beginning, such as the lower chromosphere or even the temperature minimum region. In this dense and cool plasma environment, the released energy may mostly dissipate as plasma heating due to the larger plasma beta, thus resulting in a compact BP or UV burst. As the emergence of the dipole goes on, the dipole will interact with ambient fields at a higher height and the interaction (i.e., squeezing effect) between them will be dramatically enhanced. As a result, the current sheet becomes thinner and may extend to the upper atmosphere. This can lead to two important results: (1) the thinner current sheet may introduce tearing mode instability, leading to a higher reconnection rate; (2) the extended current sheet may reach the upper chromosphere and even the transient region (with a decreased plasma density and a lower plasma beta). Therefore, the reconnection at this stage should have a more effective energy release. The abundant released energy can heat the less dense plasma more easily and simultaneously transform into plasma kinetic energy, driving the cool plasma ejection. The cold dense plasma may come from a small minifilament in at least two cases. In the end, the emerging negative flux fully cancelled with the positive-polarity ambient fields, and the plasma heating near the macrospicule bases gradually ceased. As a result, the enhanced emission progressively fades out from UV to chromospheric emission due to the cooling of plasma near the reconnection site. This conjecture interprets our current observations in a two-stage reconnection manner. 

\par
In our study, we find a positive correlation between the deceleration and peak velocity of the macrospicules, which is similar to previous results found in small-scale chromospheric jets, spicules, and dynamic fibrils fibers \citep{2004Natur.430..536D, 2006ApJ...647L..73H, 2007ApJ...655..624D,2007ApJ...666.1277H}. This suggests that the triggering of macrospicules might  involve shock waves. The origin of the shock waves has been attributed to two possible mechanisms: the leakage of $p$-mode waves from the photosphere \citep[e.g.,][]{2004Natur.430..536D, 2006ApJ...647L..73H, 2007ApJ...655..624D, 2007ApJ...666.1277H,2017ApJ...838....2Z} or the energy release in reconnection \citep[e.g.,][]{2013PASJ...65...62T, 2020PASJ...72...75K, 2014ApJ...790L...4Y, 2017ApJ...835..240S,2018ApJ...854...92T}. In our case, we tend to believe that magnetic reconnection plays a key role in driving the upward shock waves, as suggested by simulations \citep[e.g.,][]{2013ApJ...770L...3K, 2013PASJ...65...62T, 2020PASJ...72...75K}. This speculation can be supported by two observational evidence, including (1) the macrospicules are closely related to reconnection-induced base BPs, and (2) the period of the macrospicules (10 minutes) are much longer than that of the leakage of $p-$mode wave from the photosphere (3$-$5 minutes). In our sketch in Figure 3 (c), recurrent magnetic reconnection occurring in the lower atmosphere can successively generate magnetoacoustic waves. They can propagate upward and develop into shock waves (due to amplitude growth caused by the density change), eventually resulting in the observed macrospicules.

\par 
The observational features of rotational motions in jet spires were reported by previous observational and theoretical studies \citep[e.g.,][]{2009ApJ...691...61P,2011ApJ...735L..43S,2012ApJ...745..164S,2019ApJ...887..239Y,2021ApJ...911...33C}. The rotating motion in jet spires can be explained as a result of the reconnection between twisted flux ropes and/or filaments and the ambient open magnetic fields. The simultaneous blueshift and redshift on each side of the macrospicules could also be interpreted as an indication of the rotating motion \citep[e.g.,][]{1998SoPh..182..333P, 2010A&A...510L...1K, 2011A&A...532L...9C}. Similarly, some observational studies have also recorded that many spicules exhibit spinning or twisting motion during their lifetime \citep[e.g.,][]{1968SoPh....3..367B,1968SoPh....5..131P,2008ASPC..397...27S}. \citet{2020ApJ...893L..45S} conjectured that the rotational motion of spicules may be from erupting minifilaments, which transfer twist fields via reconnection. 
Based on the  earlier H$\alpha$ observations of macrospicules in a coronal hole, \citet{2005ApJ...629..572Y} claimed that a small portion of macrospicules seems to be initiated from so-called loop eruptions (or mini-filament eruptions). On the contrary, with the contemporary H$\alpha$ observations,  \citet{2000ApJ...530.1071W} proposed that macrospicules and eruptive mini-filaments should be two different forms of activity.
Fortunately, with the higher-resolution observations from NVST, in our current work, several small minifilaments are unambiguously identified at the base of recurrent macrospicules, which are associated with magnetic flux emergence and cancellation. This provides a solid clue that macrospicules should be a scaled-down version of solar jets, which are initiated with a scaled-down small minifilament eruption at their bases due to photospheric flux cancellation. Meanwhile, the presence of small minifilament eruptions at the base of macrospicules also reinforces
that the result of the rotation motion in macrospicule spire may be from the reconnection between the small minifilaments (or the twisted fields) and the ambient fields. Of course, we can not fully rule out the possibility of the coexistence of upflows and downflows in the spire of the macrospicules, since the plasma upward and downward can be observed around the end of a macrospicule but the start of the following one.

\par 
\cite{2019Sci...366..890S} revealed an association between spicules and the appearance of opposite-polarity magnetic fields at their bases. This reveals an indirect clue that spicules may be generated by magnetic reconnection between the weak opposite-polarity patch and the expanded network fields in the chromosphere, as the results found in larger-scale reconnection-driven coronal jets \citep{2016ApJ...832L...7P,2021RSPSA.47700217S}. As mentioned above, our observation provides more convincing evidence of reconnection at the base of macrospicules than their work, because macrospicules can be better spatially resolved by current solar instruments. Even though similar bright patches have not been detected at the base of spicules by current solar instruments, we tend to believe that macrospicules might be scaled-down solar jets or scaled-up versions of spicules. One important reason is that opposite-polarity fluxes are both found at the base of spicules in \cite{2019Sci...366..890S} and macrospicules in our work. In the future, higher-resolution observations are needed to confirm or deny this possibility.
In our observation, the emerged and canceled magnetic flux is of the order of  $10^{19} Mx$; the flux emergence and cancellation rates are of the order of $10^{15}\Mx$. The orders of magnitude of the cancellation rate is larger than quiet jets ($10^{14}\Mx$) and coronal jets ($10^{14}\Mx$), and similar to AR jets ($10^{15}\Mx$) \citep[e.g.,][]{2018ApJ...864...68S}. However, the canceled flux and cancellation rates are 3 and 2 orders of magnitude larger than the spicules reported in \citet{2019Sci...366..890S}, respectively. These differences are reasonable because the launch of larger-scale macrospicules needs more magnetic energy than spicules. However, whether the triggering process of spicules is fully identical to that of macrospicules as observed here needs further study.

\par

%%%%%%%%%%%%%%%%%%%%%%%%%%%

\acknowledgments
The authors would like to thank the data provided by the {\em NVST} and {\em SDO} teams. This work is supported by the Natural Science Foundation of China (12173083, 12103005, 11922307, 11773068), the Yunnan Science Foundation for Distinguished Young Scholars (202101AV070004), the National Key R\&D Program of China (2019YFA0405000), the Specialized Research Fund for State Key Laboratories, and the Yunnan Key Laboratory of Solar Physics and Space Science (202205AG070009).

\begin{figure}[t]    %%%%%%%%%%%%%%%%%% FIGURE 1
\centerline{\includegraphics[width=0.75\textwidth,clip=]{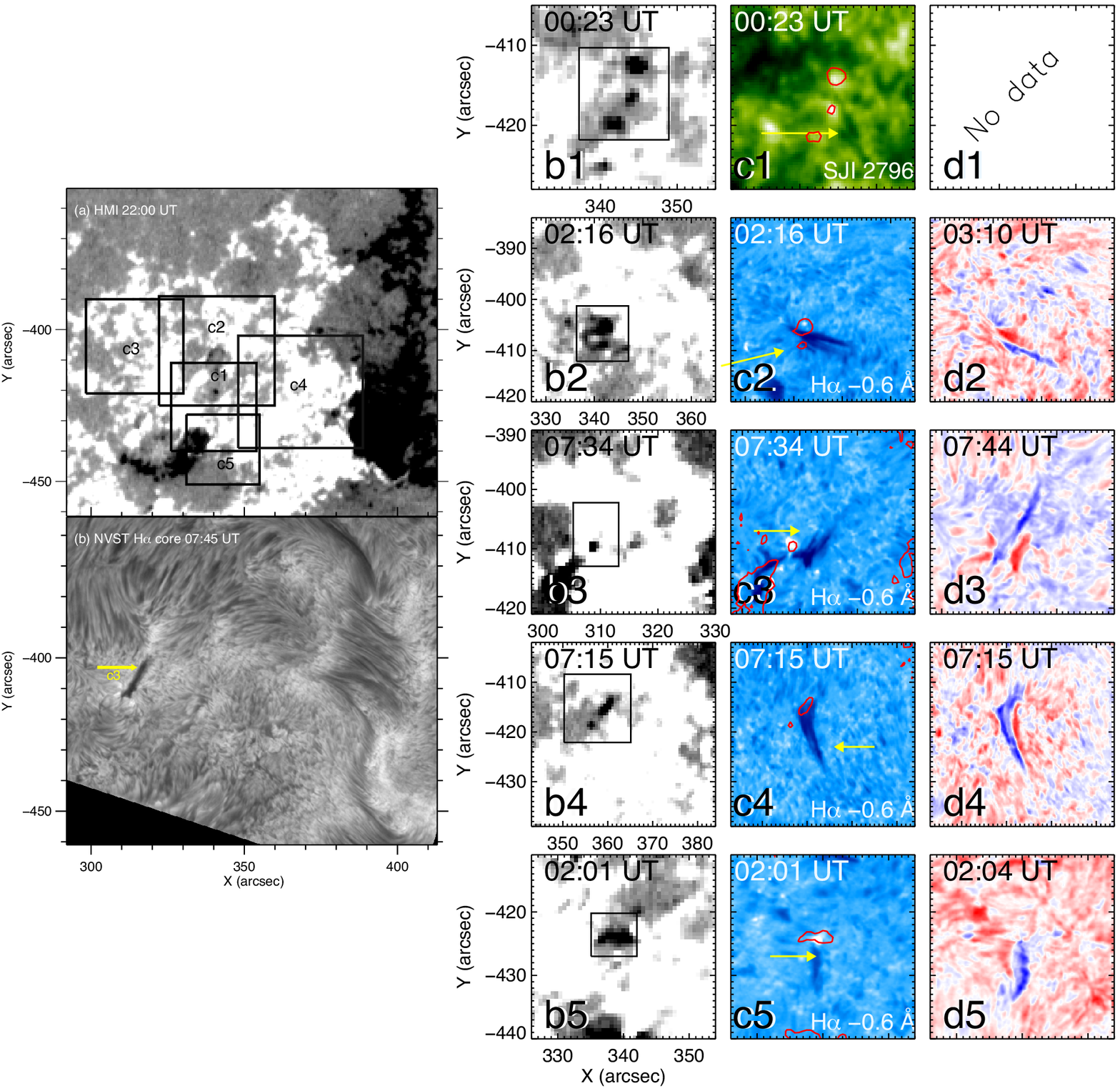}}
\caption{Panel (a): HMI light-of-sight magnetogram is taken at 22:00 UT on November 10, 2020. The overlaid five boxes labeled as c1$-$c5 mark the field of view (FOV) of the macrospicules studied here in panels (c1)$-$(c5). SJI 2796 \AA\ image (c1) and NVST Ha -0.6 \AA\ images (c2)$-$(c5) display the five group recurrent macrospicules. The HMI LOS magnetograms (b1)$-$(b5) correspond to the same time in the panel (c1)$-$(c5) images. The yellow arrows denote the macrospicules under study. Panels (d2)$-$(d5): The constructed Doppler maps show the blueshift and redshift emissions on each side of the macrospicules. Macrospicules as the absorption observation features on disk, Doppler blueshift (or redshift) corresponds to the negative (or positive) proxy index signal. An online animation of the five groups of recurrent macrospicules is available. The red contours are the negative field that is embedded in a large patch of positive flux in the panels (c1)$-$(c5) and in the animation. The animation duration is 3s.}
\label{fig1}
\end{figure}

\begin{figure}[b]    %%%%%%%%%%%%%%%%%% FIGURE 2
\centerline{\includegraphics[width=0.7\textwidth, clip=]{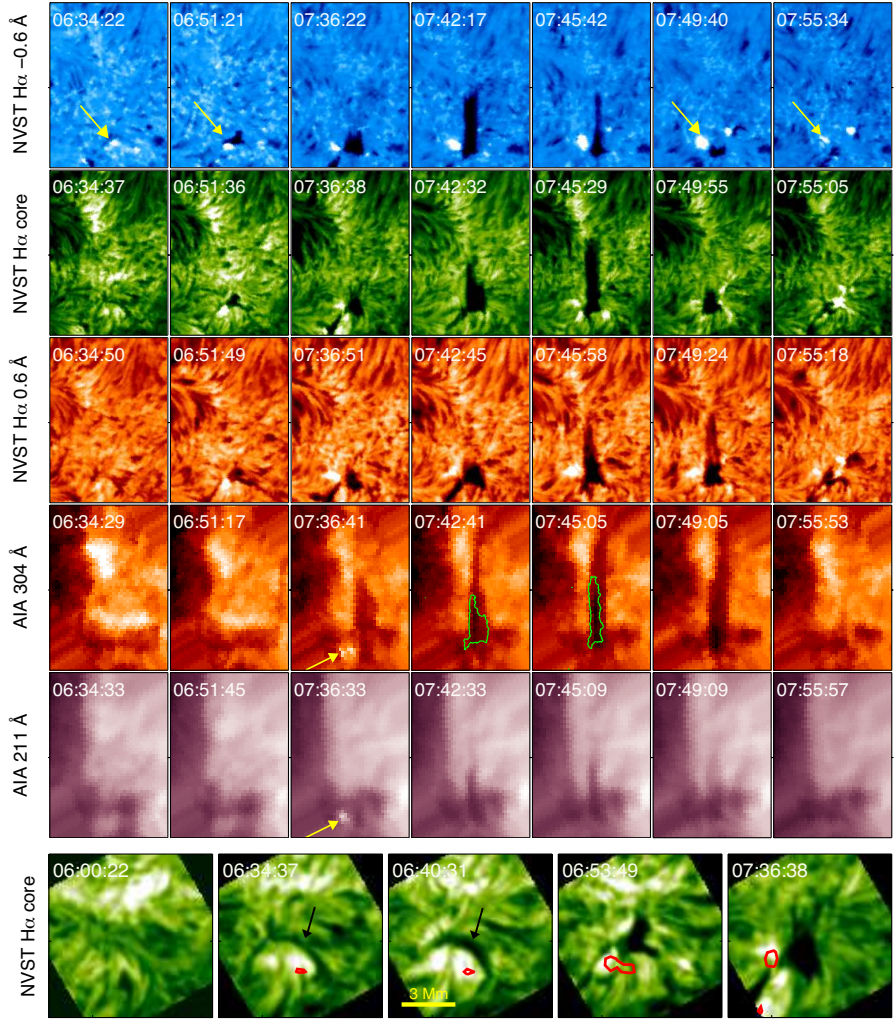}}
\caption{Dynamics of the macrospicule in the region Box c3 in Figure 1 are observed in NVST H$\alpha$.  NVST Ha core, wing (-0.6 and +0.6 ), AIA 304, and 211 \AA\ images present the temporal evolution of the macrospicule. The yellow arrows denote the BPs, and the green contours outline the macrospicule from the H$\alpha$ core images for the corresponding time. The FOV of NVST and SDO images are 25\arcsec $\times$ 35 \arcsec, while in the last row, H$\alpha$ core images are 15\arcsec $\times$ 15 \arcsec, which corresponds to the base region of the macrospicule. The black arrows point out the small minifilament. The red contours mark the negative magnetic field and the corresponding scale levels are -10 G. This negative-polarity flux are fully surrounded by an ambient positive-polarity flux (also see Figure 1 (b3)). We rotated all the images from row 1 to row 6 by 30 degrees to the north.
}
\label{fig2}
\end{figure}

\par
\begin{figure}[b]    %%%%%%%%%%%%%%%%%% FIGURE 3
\centerline{\includegraphics[width=0.8\textwidth,clip=]{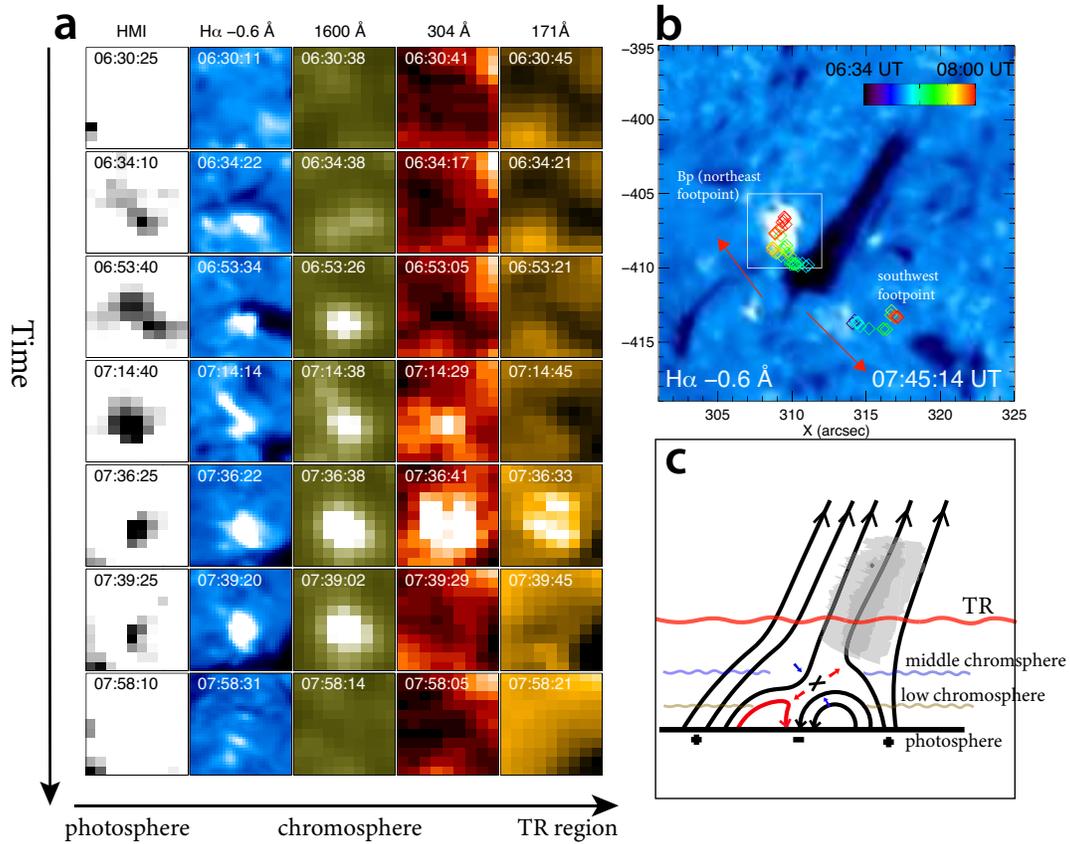}}
\caption{Panel (a): The BP appearing at the jet base of a macrospicule was observed by HMI LOS magnetogram, NVST H$\alpha$, AIA 1600 \AA, 304 \AA~, and 171 \AA. The small field of view tracking the BP is $5''\times 5''$, which is shown in the white box in panel (b). The first column displays the process of the negative field emerging in the large patch of positive flux. The scale levels of positive flux and negative flux are $\pm$ 50 G. Panel (b) shows a representative macrospicule, whose FOV is $24''\times 24''$. The centroids of two footpoints (BP and southwest footpoint) are shown in panel (b) as diamonds from black to red, delineating the trajectory of them moving with time. The red arrows denote the direction of spreading of the two footpoints. And Panel (c) plots a schematic illustration of the representative macrospicules. The red and black curves with arrows represent magnetic field lines. The black cross indicates the reconnection position. The shaded region on the open field lines represents the macrospicule spire that formed from the material of the small minifilament when it erupted. An animation of the panel (b) is available, showing the spreading of the two footpoints from 06:29 UT to 08:00 UT. The animation duration 2s.}

\label{fig3}
\end{figure}

\begin{figure}    %%%%%%%%%%%%%%%%%% FIGURE 4
\centerline{\includegraphics[width=0.75\textwidth,clip=]{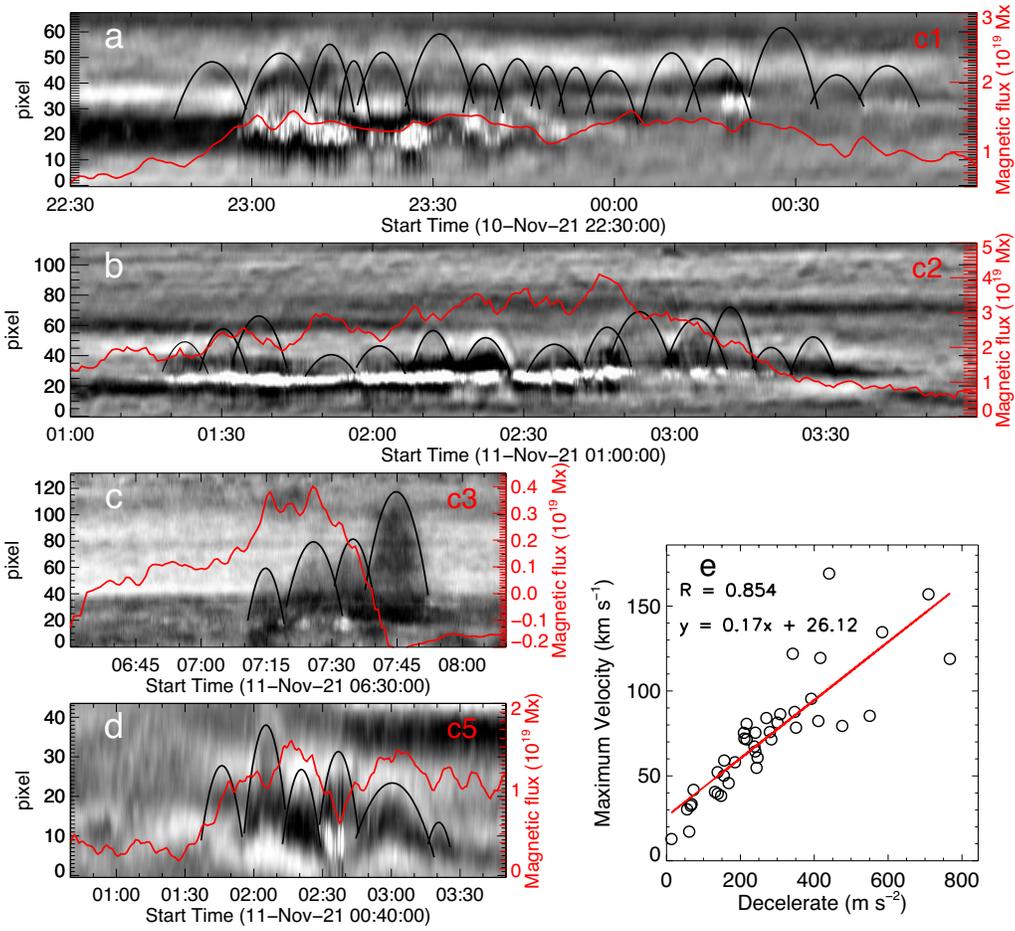}}
\caption{The dynamics of macrospicules. Panels (a)$-$(d) show four examples of dynamic macrospicule , in which the red curve lines represent each negative magnetic flux variation that are indicated in the black box in Figure 1 (b1)$-$(b5). The time-distance images from AIA 193 channel are plotted along the each group macrospicules' main axis. Panel (e) is a scatter plot of the relationship between the deceleration and maximum velocity, which is measured from a parabolic fitting of the trajectories of the macrospicules front as shown in the left column.
}
\label{fig4}
\end{figure}

\begin{figure}    %%%%%%%%%%%%%%%%%% FIGURE 5
\centerline{\includegraphics[width=0.8\textwidth,clip=]{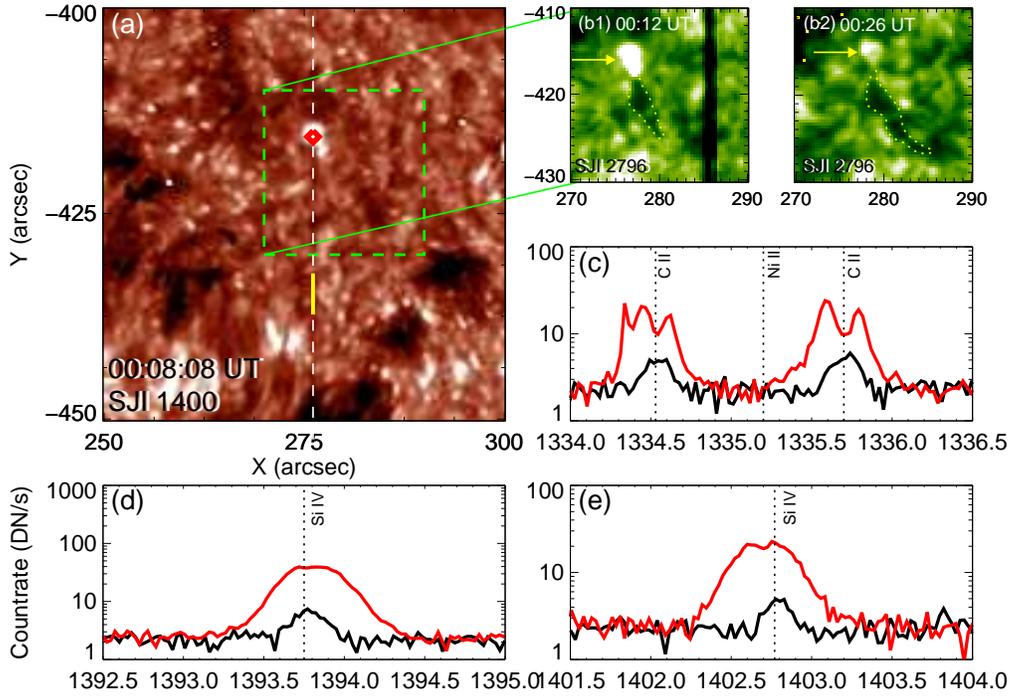}}
\caption{Panel (a): IRIS 1400 \AA\ slit-jaw image is taken at 00:08:08 UT on November 11, 2020. The slit location is shown in the white dashed line, and the green box indicates the FOV of the panels (b1$-$(b2). The yellow dashed lines outline the macrospicule in (b1)$-$(b2) 2796 \AA\ images at different times, and the yellow arrows denote the BP at the base of the macrospicule. Panels (c)$-$(e): The red lines represent the spectra at the location indicated by the red diamond in panel (a). The black lines display the reference spectra averaged over the line profiles within the section indicated by the yellow line in panel (a).}
\label{fig5}
\end{figure}

%{\tablenotemark{a}} 
% \begin{table*}[]
%\caption{Detailed information of the IRIS raster scans and supersonic downflows.}
%\label{tab1}
%\resizebox{\textwidth}{!}{
\linespread{1.1}
\begin{table*}[htb]\scriptsize %\tiny
\caption{Information about all Macrospicule Events} \label{t1}
\setlength{\tabcolsep}{0.6mm}{
\begin{tabular}{clcclccccccc}
\hline
\hline
Event & Telescope/ & Date\tablenotemark{a}/   & Corresponding\tablenotemark{a}    & Maximum \tablenotemark{b}        & Velocity\tablenotemark{c}      & Deceleration\tablenotemark{c}  & BP \tablenotemark{d} &Phtospheric & Small & Emergence/ \\
   ID      &    Passbands       & Time       & AIA Date/         & Length/     &       &        & Duration   & Flux  &minifilament  &cancellation\\
         &           & (UT)        & Time (UT)            &  Width (Mm)  & ($\speed$)       & ($\accel$)       &  Time (UT)   & Features & signal&rate($10^{15}\Mx$ )       \\\hline
C1  & IRIS      & 00:03-00:33  & 23:00-00:36    & 6.24 /            & 38.25-118.9  & 138.4-767.3  &      22:02-00:59  &   emergence/  & Unclear & 5.747/2.38\\
         &     2796 \AA\       &  &       &       1.31                 &              &    & &cancellation    \\
         &     1400 \AA\       &  &       &   &              &                        &              &                                       & \\ \hline
C2   & NVST      & 01:46-03:30 & 01:18-03:30        & 14.8/              & 40.56-119.45 & 131.2-416.5  &    00:43-03:40     &  emergence/  & Unclear &  4.245/10.76  \\
         &   H$\alpha$ core         &  &       &     1.62        &                      &              &                                       & cancellation \\
         &    H$\alpha$ $\pm$ 0.6 \AA\       &  &    &                         &              &              &                                       &  \\ \hline
C3   & NVST      & 06:51-07:53 & 07:19-07:55      & 8.49/             & 121.95-169.17 & 341.8-709.4  &       06:40-08:00  &   emergence/  &   Yes & 1.407/4.271\\
         &   H$\alpha$ core         &  &       &     1.85     &                        &              &                                       & cancellation \\
         &    H$\alpha$ $\pm$ 0.6 \AA\       &  &  &                         &              &              &                                       &  \\ \hline
C4   & NVST      & 06:50-08:46 & 06:55-08:48         & 20.32/             & \dots  \tablenotemark{e}            & \dots  \tablenotemark{e}          &      06:42-08:59        &   emergence/  &   Yes  &  4.81/5.56   \\
         &   H$\alpha$ core         &  &       &     5.66                   &              &              &                                       & cancellation \\
         &    H$\alpha$ $\pm$ 0.6 \AA\       &  &    &                         &              &              &                                       &  \\ \hline
C5   & NVST      & 01:56-03:16 & 01:59-03:20       & 8.34               & 12.82-41.77  & 13.5-73.2    &          01:55-04:00        &   emergence/     &   Unclear     &4.761/7.57         \\
         &   H$\alpha$ core         &  &       &    2.18                   &              &              &                                       & cancellation \\
         &    H$\alpha$ $\pm$ 0.6 \AA\       &  &    &                       &              &              &                                       &  \\\hline
%\multicolumn{10}{l}{BBBBB}   
\end{tabular}
}
\\
\footnotesize{\textbf{Note.} The observed dates are all from 2020 November 11, except that the AIA corresponds to event 1 begins on November 10. }\\
\footnotesize{$^a$ The Date/Time and Corresponding AIA Date/Time represent the obtained time of macrospicules in each Telescope. }
\\
\footnotesize{$^b$ The Maximum Length and Maximum Width are from the NVST H$\alpha$ core (or IRIS 2796 \AA\ ).} 
\\
\footnotesize{$^c$ The Velocity and Deceleration come from a parabolic fitting of the trajectories of the macrospicules front as displayed in Figure 4.}
\\
\footnotesize{$^d$ The BPs duration times at the base of macrospicules are from AIA 1600\AA~, since this 1600\AA~ channel can provide a more persistent observation.}
\\
\footnotesize{$^e$ The reason for the absence of measured values is that the velocity and acceleration of the macrospicule cannot be determined because of its sliding behavior.}
\\

\end{table*}

\end{document}